\documentclass[aps,nofootinbib,pra,notitlepage,twocolumn]{revtex4-1}
\usepackage{amsfonts,amsmath,amssymb,amsthm}
 \usepackage{array,bm,color}
\usepackage{epsfig,graphicx,nomencl,revsymb4-1,upgreek,url}
\usepackage{hyperref}
\usepackage{algorithm}
\usepackage{algpseudocode}
\usepackage{graphicx}
\usepackage{calc}
\usepackage{siunitx}
\usepackage{tabularx}
\newcolumntype{Y}{>{\centering\arraybackslash}X}
\graphicspath{{./figures/}}
\hypersetup{colorlinks=true, pdfauthor=Anthony M. Polloreno, pdftitle=Robustly decorrelating errors with mixed quantum gates, citecolor=blue, linkcolor=blue}
\newcommand{\tr}{{\rm Tr\thinspace}}

\newcommand{\braket}[2]{\left\langle #1 | #2 \right\rangle}

\newcommand{\expect}[1]{\ensuremath{\left\langle{#1}\right\rangle}}

\newcommand{\order}[1]{\mathcal{O}\left( #1 \right)}

\newcommand{\note}[1]{}

\newcommand{\actual}{\ensuremath{\tilde{\mathsf{G}}}}
\newcommand{\actmat}{\ensuremath{\tilde{\mathcal{G}}}}
\newcommand{\target}{\ensuremath{{\mathsf{G}}}}
\newcommand{\tarmat}{\ensuremath{{\mathcal{G}}}}
\newcommand{\error}{\ensuremath{{\mathsf{E}}}}
\newcommand{\errmat}{\ensuremath{{\mathcal{E}}}}
\newcommand{\genmat}{\ensuremath{{\mathcal{L}}}}
\newcommand{\vectorize}[1]{\ensuremath{\mathsf{vec}\left(#1\right)}}
\newcommand{\AGI}{\ensuremath{\epsilon_\mathcal{F}}}
\newcommand{\dnorm}{\ensuremath{\epsilon_\diamond}}
\newcommand{\identmat}{\ensuremath{\mathcal{I}}}
\newcommand{\ident}{\ensuremath{\mathsf{I}}}
\newcommand{\0}{\ensuremath{\mathbf{0}}}
\newcommand{\weight}{\ensuremath{w}}

\newcommand{\bunderbrace}[2]{
  \begin{array}[t]{@{}c@{}}
  	#1\\
  	\parbox{\widthof{#1}}{$\scriptscriptstyle#2$}
  \end{array}}

\begin{document}
\title{Robustly decorrelating errors with mixed quantum gates}

\author{Anthony M. Polloreno}
\email[Email: ]{ampolloreno@gmail.com}
\altaffiliation[Current address: ]{University of Colorado, Boulder, CO}
\affiliation{Rigetti Computing, 2919 Seventh Street, Berkeley, CA, United States of America}
\author{Kevin C. Young}
\affiliation{Quantum Performance Laboratory, Sandia National Laboratories, Livermore, CA, United States of America}

\date{\today}

\begin{abstract}
\noindent Coherent errors in quantum operations are ubiquitous. Whether arising from spurious environmental couplings or errors in control fields, such errors can accumulate rapidly and degrade the performance of a quantum circuit significantly more than an average gate fidelity may indicate. As Hastings \cite{Hastings2017} and Campbell \cite{Campbell2017} have recently shown, by replacing the deterministic implementation of a quantum gate with a randomized ensemble of implementations, one can dramatically suppress coherent errors. Our work begins by reformulating the results of Hastings and Campbell as a quantum optimal control problem. We then discuss a family of convex programs able to solve this problem, as well as a set of secondary objectives designed to improve the performance, implementability, and robustness of the resulting mixed quantum gates. Finally, we implement these mixed quantum gates on a superconducting qubit and discuss randomized benchmarking results consistent with a marked reduction in the coherent error.\\
{[1]} M. B. Hastings, \emph{Quantum Information \& Computation} 17, 488 (2017). \\
{[2]} E. Campbell, \emph{Physical Review A} 95, 042306 (2017).
\end{abstract}

\pacs{}

\maketitle

\section{Introduction}
\label{sec:introduction}
\noindent The ultimate impact of a gate error on the performance of a quantum circuit depends strongly on both the magnitude and the nature of the error. Systematic, or \emph{coherent}, errors can arise from poorly calibrated controls or approximate gate compilations that induce repeatable, undesired unitary errors on the state of a quantum information processor. Errors of this type are correlated in time and may add up constructively or destructively, depending on details of the circuit in which they appear. This can make it difficult to construct tight analytic bounds on circuit performance \cite{Beale2018}, and numerical studies are often limited by the high computational cost of modeling coherent errors. Contrast this against random, or \emph{stochastic}, errors which often result from high-frequency noise in the controls or the environment. Systems with stochastic errors can usually be accurately modeled by defining a rate of various discrete errors in the system, such as a bit flips or phase flips. These errors are significantly easier to simulate on a classical computer, and their impact on quantum circuits is much easier to estimate \cite{Beale2018}.

\begin{figure}[t]
  \centering
  \includegraphics[width=\columnwidth]{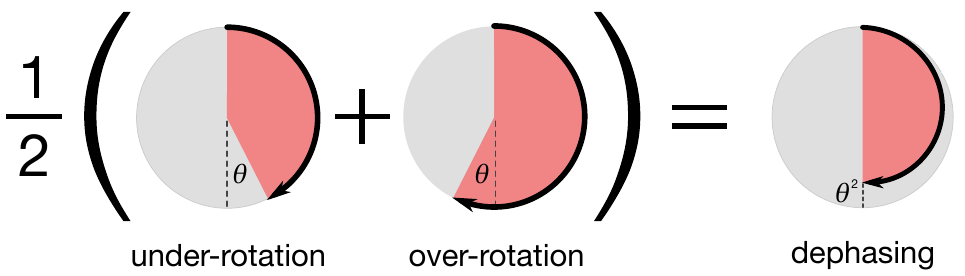}
  \caption{\textbf{Simple example of a mixed quantum gate:} Using optimal control, two implementations of a $Z_\pi$ gate are designed to have equal and opposite coherent errors (if one implementation over-rotates by a small angle $\theta$, then the other \emph{under}-rotates by $\theta$). Each time the gate is used, one of these two implementations is chosen at random. The resulting effective quantum gate is equivalent to a perfect implementation of the gate followed by dephasing with associated probability $\order{\theta^2}$.}
  \label{fig:simple_example}
\end{figure}

Campbell \cite{Campbell2017,Campbell2019} and Hastings \cite{Hastings2017} have developed a technique for suppressing coherent noise by replacing deterministic gate implementations with a \emph{mixed quantum gate} (MQG) consisting of a randomly sampled ensemble of implementations. They focus on errors in gate compilation, such as those that arise from the Solovey-Kitaev algorithm, for which any lingering approximation errors are generally coherent, even if the underlying native gates are perfect. Different approximate gate compilations of the same target unitary will almost certainly result in different unitary errors. So by selecting from these various compilations at random, the resulting quantum channel becomes a mixture of unitaries \cite{DBLP:journals/corr/cs-CC-0012017}, which can have significantly less coherent error than any single compilation on its own. A very simple example of this reduction in coherent error by mixed quantum gates is illustrated in Fig.~\ref{fig:simple_example}. 

Randomized protocols have a long history of outperforming their deterministic counterparts\cite{Viola2005, Santos2006}. In circuit models, Pauli Frame Randomization \cite{Wallman2016, 1803.01818, dahlen2018experimental} has been shown to reduce coherent errors by introducing randomness in a compilation step. MQGs are an alternative approach to solving this problem. Because they are implemented by randomly drawing from different gate realizations, they only require randomness at run-time, allowing them to be easily implemented in hardware. Moreover, when the construction of MQGs is posed as a numerical optimization problem, numerous useful modifications can be made to augment their performance.

In this article, we extend the work of Campbell and Hastings. These prior works are focused on removing coherent errors that arise from compiled (e.g,, via Solovay-Kitaev algorithm) quantum gates. Errors arising from approximate compilation are unitary, and this previous work illustrates how appropriate averaging over compilations can transform these coherent errors into incoherent (stochastic) errors. In this work, we focus on a more general class of errors arising in the context of optimal control. Furthermore, we construct a series of explicit convex optimization problems whose solutions consist of distributions over gate implementations and whose effective errors can be sculpted to suit various experimental or theoretical needs. This includes reduction of a gate’s diamond error, elimination of coherent errors, elimination of off-axis stochastic errors, and the inclusion of robustness to drift and model uncertainty.  We demonstrate our results on a superconducting testbed device at Rigetti Computing. In a simple experiment based on randomized benchmarking circuits, MQGs demonstrate a marked improvement in error rates and a reduced variance in circuit outcome probabilities, consistent with a significant reduction in the coherence of the error\cite{Ball2016}. We further apply our methods in simulation, constructing both single- and two-qubit mixed unitary controls that are robust to drift and uncertainty in the control parameters. These robust MQGs are insensitive to calibration errors, but do not require the temporal overhead of robust coherent control pulses \cite{1203.6392}.

\section{Preliminaries}
\label{sec:preliminaries}

\subsection{Representing Errors in Quantum Gates}
\label{sec:rep_errors}
\noindent In order to implement a desired unitary gate, $\target$, on a quantum device, one generally applies a carefully tuned sequence of classical control fields.  But fluctuations in the environment or imperfections in these controls can cause the state of the qubits to change in a way that is different from what was intended, \emph{i.e.}, there are errors in the gate. If the device is fairly stable with time\cite{1907.13608} and context\cite{Rudinger2019}, then we can accurately model the gate action using a completely positive, trace-preserving (CPTP) map, $\actual$ acting on the Hilbert space of the target qubits. This map can always be written as $\actual = \error\circ\target$, where $\error = \actual\circ\target^{-1}$ is the \emph{error map}, which is itself CPTP because $\target$ is unitary.

CPTP maps possess a number of useful representations, including Kraus operators\cite{1983}, Choi matrices\cite{Choi1975}, and Jamiolkowsi states\cite{yczkowski2004}. But for the purposes of this article, the \emph{process matrix} representation will be particularly convenient\cite{OBrien2004}, and we shall denote the process matrix associated with a given CPTP map $\target$ with its corresponding calligraphic character, $\tarmat$. For a $d$-dimensional quantum state, the process matrix is a $d^2\times d^2$ matrix. A key feature of process matrices is that they are composable and act through the usual matrix multiplication on the vectorized quantum state:
\begin{align}
	\vectorize{\actual(\rho)}
		&= \actmat\cdot\vectorize{\rho} \\
		&= \errmat\cdot\tarmat\cdot\vectorize{\rho}
\end{align}
The $\mathsf{vec}$ operation is typically performed using a basis of matrix units, for which $\vectorize{\rho}$ would be the column vector obtained by stacking the columns of $\rho$. In this work, however, we shall use a basis of Pauli matrices, defining $P = \left\{I, \sigma_x, \sigma_y, \sigma_z\right\}^{\otimes n}$ as the collection of all $4^n$ $n$-qubit Pauli operators (including the identity).  In this basis,
\begin{equation}
  \vectorize{\rho}_i = \expect{P_i} = \tr\!\left(P_i\; \rho \right),
\end{equation}
and
\begin{equation}
	(\actmat)_{ij} = \frac{1}{d}\tr\!\left(P_i \;\actual\!\left(P_j\right) \right).
\end{equation}
Thus, as an example, we can consider a coherent $X$ rotation of angle $\theta$ (denoted $X(\theta)$) on a single qubit that has the following action on the expectation values:
\begin{eqnarray}
    \expect{Z} &\to& \cos{(\theta)}\expect{Z} + \sin{(\theta)}\expect{Y}\\
    \expect{Y} &\to& \cos{(\theta)}\expect{Y} - \sin{(\theta)}\expect{Z}\\
    \expect{X} &\to& \expect{X}
\end{eqnarray}
The matrix representation is then given as 
\begin{equation}
X(\theta) =
	\left(\begin{array}{ccccc}
		1 & 0 & 0 & 0 \\ 
		0 & 1 & 0  & 0  \\
		0 & 0 & \cos{(\theta)} & \sin{(\theta)}\\
		0 & 0 & -\sin{(\theta)} & \cos{(\theta)}
	\end{array} 	
	\right)
\end{equation}
Contrast this with a stochastic error, $SX(p): \rho\to (1-p)\rho + pX\rho X$. This is given by
\begin{equation}
SX(p) =
	\left(\begin{array}{ccccc}
		1 & 0 & 0 & 0 \\ 
		0 & 1 & 0  & 0  \\
		0 & 0 & 1-p & 0\\
		0 & 0 & 0 & 1-p
	\end{array} 	
	\right)
\end{equation}
Both of these channels are seen to have the vector $\{1,0,0,0\}$ in the first column - this is a property of all unital channels.

More generally, process matrices written in this Pauli basis are often referred to as \emph{Pauli transfer matrices}\cite{Chow2012}. Error maps represented in this basis take the particularly nice form,
\begin{equation}\label{eq:process_matrix}
\errmat =
	\left(\begin{array}{c|cccc}
		1 &  & \vec{0}^T & \\ 
		\hline & &  &  \\
		\vec{m} &  & R &  \\
		 &  &  & 
	\end{array} 	
	\right)
\end{equation}
The top row of all trace-preserving (TP) error maps is fixed to $\{1,0,0,0,\cdots\}$ and the remainder of the first column, $\vec{m}$, describes any deviations from unitality, as might arise from amplitude damping \cite{preskill1997lecture}. If the error map is unitary, then the error is called \textit{coherent} and the unital submatrix $R$ is a rotation matrix.  Importantly, if $R$ is diagonal, then the error is Pauli-stochastic, with each diagonal entry corresponding to the probability that its associated Pauli error occurs in each application of the gate.

In general an error will neither be fully coherent nor Pauli-stochastic and so it is useful to define what we mean by the coherence of arbitrary maps\cite{CarignanDugas2019}.
 One way to measure this is with a quantity called the unitarity \cite{Wallman2015}:
\begin{align}
u(\mathsf{E}) = \frac{d}{d-1}\int d\psi ||\mathbf{n}[\mathsf{E}(\psi)] - \mathbf{n}[\mathsf{E}(\identmat_d/2)||^2
\end{align}
where $\mathbf{n}(\rho)$ is the Bloch vector of the state $\rho$, and the integral is over all $d$-dimensional pure states using the Haar measure. Yet another way of quantitatively describing how unitary a map is with protocols that assess how errors accumulate on different input states\cite{1907.05950}. In this work, we will only discuss the relative unitarity of channels, and therefore are content if the unitarity of our error map decreases by any of these metrics.

In what follows it will be useful to define the \emph{error generator}, $\genmat$, associated with a faulty gate: 
\begin{align}
	\errmat 
		&= \exp\left(\genmat\right) \\
	\label{eq:generator}
		&= \identmat_d + \genmat + \frac{1}{2}\genmat^2 + \order{\genmat^3}.
\end{align}
If an implemented gate is relatively close to its target, then the error generator will be small under any of the usual matrix norms, and the Taylor expansion above may reliably be truncated at first or second order. Additionally, if the errors are unitary, then the error generator will include Hamiltonian terms\cite{2103.01928}.

\subsection{Mixed Quantum Gates}
\label{sec:mqg}
\noindent Suppose that we have access to an ensemble of distinct implementations, 
$\{\actual_i\}$, of a target gate, $\target$.
Each time the gate is to be applied to the system, we randomly select an implementation from this ensemble such that the probability of drawing $\actual_i$ is $\weight_i$ (and we ensure that $\sum_i \weight_i=1$). This procedure is operationally indistinguishable from always applying the effective channel, 
\begin{align}
	\actual_{\rm eff} &= \sum_i \weight_i \actual_i = \left(\sum_i \weight_i \error_i \right) \circ\target
\end{align}
We call such randomized quantum operations \emph{mixed quantum gates} or MQGs. Error metrics for these MQGs can be computed in terms of their effective error map, 
\begin{equation}
	\label{eq:effective_error}
	\error_{\rm eff} = \sum_i \weight_i \error_i.
\end{equation}
Two important error metrics are the average gate infidelity (AGI), $\AGI$\cite{Johnston2011}, and the diamond distance to the target (or simply, the \emph{diamond distance}), $\dnorm$ \cite{watrous2018theory}, defined as:
\begin{align}
	\label{eq:def_agi}
	&\AGI(\error) = \frac{d^2 - \tr{\errmat}}{d^2 + d},\\
	\label{eq:def_diamond}
	&\dnorm \left(\error\right)
		= \frac{1}{2} \sup_\rho \vert \vert (\ident_d\otimes \ident_d)(\rho) 
										  - (\error \otimes \ident_d)(\rho) \vert\vert_1,
\end{align}
where $d=2^n$, with $n$ being the number of qubits, and $\ident_d$ is the $d$-dimensional identity operator.  In Eq.~\eqref{eq:def_agi}, we have written the AGI for an error map, $\error$, in terms of its associated Pauli transfer matrix, $\errmat$. If the error channel is purely stochastic, then $\dnorm(\error) = \AGI(\error)$, but if the error channel has a unitary component, then the diamond distance will generically be larger than the average gate infidelity \cite{1511.00727}. The diamond distance is subadditive \cite{watrous2018theory} under gate composition, so is particularly useful for constructing bounds on quantum circuit performance: the total variation distance between the outcome probabilities of a faulty and perfect quantum circuit is less than or equal to the sum of the diamond distances for gates that compose the circuit \cite{aharonov1998quantum}.

Because $\AGI$ is linear in the error map, we have:
\begin{equation}
	\AGI\!\left(\error_{\rm eff}\right) = \sum_i \weight_i \,\AGI(\error_i).
\end{equation}
That is, the AGI of the effective error channel is simply the weighted average of the component AGIs, so MQGs provide little benefit for reducing the AGI. However, the diamond distance is a nonlinear function of error map. As we show in the appendix (\ref{sub:diamond_distance_inequality}),
\begin{equation}
	\label{eq:diamond_ineq}
	\dnorm\!\left(\error_{\rm eff}\right) \le \sum_i \weight_i \, \dnorm(\error_i).
\end{equation}
So by mixing various implementations, each with a different error channel, the resulting channel can have a diamond distance error that is less than the diamond distance error of any of the component implementations. 

\begin{figure}
  \centering
  \includegraphics[width=\columnwidth]{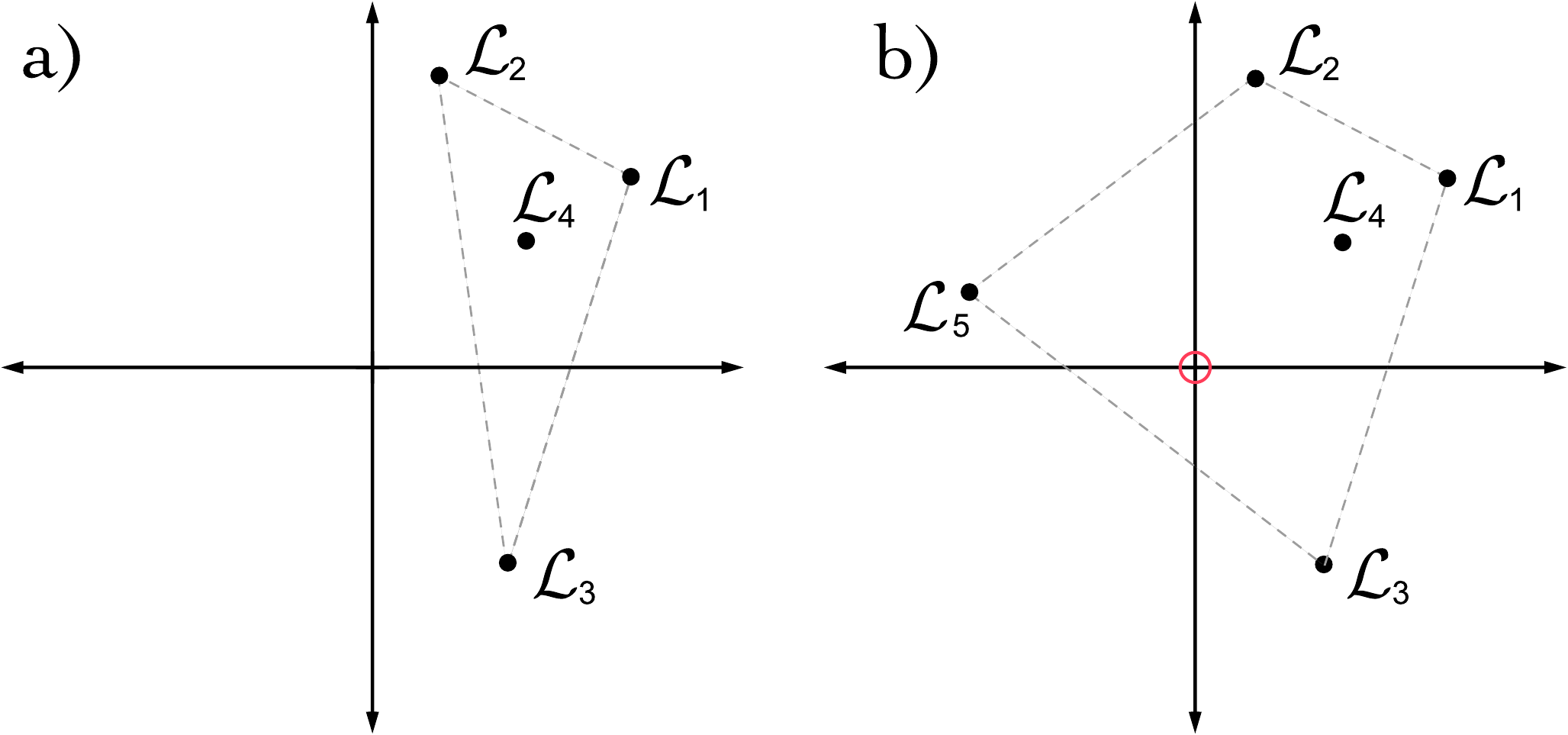}
  \caption{\textbf{Geometric condition for existence:} A target unitary gate can be implemented a number of ways, each associated with a different error generator, $\genmat_i$. These generators lie in a vector space, which we illustrate as two-dimensional here. a) Four error generators. The origin is not contained in their convex hull, so there are no generator-exact MQGs. b) After including an additional control solution, the convex hull grows to contain the origin, and so a generator-exact MQG exists. }
  \label{fig:vectorspace}
\end{figure}

Campbell \cite{Campbell2017} and Hastings \cite{1612.01011} independently considered these mixed channels using gates constructed with the Solovey-Kitaev algorithm, for which many approximate gate compilations are possible. Campbell showed that, if the error generators of some ensemble of gate compilations form a convex set containing the origin, then one can construct a MQG with quadratically suppressed diamond distance to the target. Explicitly, the weights are chosen such that the error generator is canceled to first order. Using (\ref{eq:effective_error}) and (\ref{eq:generator}), the effective error map for a MQG in terms of the component error generators is,
\begin{align}
	\label{eq:generators_and_effective_error}
	\errmat_{\rm eff} \simeq \sum_i \weight_i \left(\identmat_d + \genmat_i + \frac{1}{2}  \genmat_i^2\right)
\end{align}
These error generators are linear operators and so are elements of a vector space. If, as Campbell required, the origin lies in their convex hull (see Fig.~\ref{fig:vectorspace}), then there exists a choice for the weights, $\weight_i^*$, such that $\sum_i \weight^*=1$ and,
\begin{align}
	\label{eq:cancel}
	&\sum_i \weight_i^* \genmat_i= 0
\end{align}
When this condition is satisfied, we call the resulting operation a \emph{generator-exact} MQG. 

If the diamond distance error rates of the component gates are bounded by $\alpha$, then Campbell shows that this first-order cancellation of the error generators can ensure that the diamond distance error rate of the MQG is bounded by $\alpha^2$. For gate compilation errors, the error channels are dominated by unitary approximation error, so the error generators are Hamiltonians. Hamiltonian generators can have positive or negative coefficients, so it is possible that the origin may lie in their convex hull. If, however, the error map contains stochastic components, then the condition of Eq.~\eqref{eq:cancel} might be impossible to satisfy. Such errors always have strictly positive probabilities, so the origin can \emph{never} lie in their convex hull. This formalism can be adapted to such cases by restricting the sum in Eq.~\eqref{eq:cancel} to only the Hamiltonian component of the generators, or by replacing the exact condition with a minimization, as we discuss in Sec.~\ref{sec:mixed_unitary_processes}. 

While generator-exact MQGs can provide a guaranteed suppression of the diamond distance, their effective error channels are unlikely to be purely stochastic. Second- and higher-order terms in \eqref{eq:generators_and_effective_error} can easily contribute to lingering coherent errors that impact the efficient simulability of the channel. In order to definitively eliminate these coherent errors, we could instead seek weights, $\weight^P_i$, that annihilate the off-diagonal entries of the effective error map. The resulting diagonal error map will be Pauli-stochastic. For a single qubit gate, it takes the form,

\begin{equation}
	\errmat_{\rm{eff}} = \sum_i \weight_i^P \errmat_i = 
	\left(\begin{array}{@{}ccccc@{}}
		1 & 0 & 0 & 0 \\ 
    	0 &  p_x & 0 & 0 \\
		0 & 0 &  p_y & 0\\
		0 & 0 & 0 &  p_z\\
	\end{array} 	
	\right),
\end{equation}
where $p_x, p_y$ and $p_z$ are the rates of Pauli $X$, $Y$, and $Z$ errors, respectively. We refer to such channels as \emph{Pauli-exact} MQGs, and the same geometric argument we used in constructing generator-exact MQGs is again applicable here. The off-diagonal elements of the error map form a vector space, so we need only check that the origin is in the convex hull of the vectors of off-diagonal elements for each of the component operations, $\errmat_i$. If the error map contains non-unital components or correlated Pauli-stochastic errors (such as $X+Y$), then this strict geometric condition may not hold and no Pauli-exact MQG would exist. However, MQGs that approximate the Pauli-exact condition can be found via the convex programing techniques discussed in Sec.~\ref{sub:off_diagonals}.

\section{Constructing mixed quantum gates}
\label{sec:mixed_unitary_processes}
\noindent We now present a methodology for constructing mixed quantum channels, formalizing the intuitive approach discussed above with a series of explicit convex programs. As mentioned, our method requires two steps. The first is a control synthesis step, in which we construct an ensemble of gate implementations. Campbell and Hastings draw their ensembles from various different gate compilations, but for this work we utilize quantum optimal control theory. Many standard control-generation algorithms, such as GRAPE \cite{Khaneja2005}, take a random initial guess for the control and iteratively improve it to yield a gate that well-approximates a target. By seeding such algorithms with many initial guesses, one can quickly construct a large ensemble of approximate quantum gates, each with a different error channel. We discuss this approach in some detail in Sec.~\ref{sec:numerical_results}. 

The remainder of this section will assume such a capability for generating a large ensemble of gate implementations, and will focus instead on the second step: constructing a distribution over that ensemble that reduces the coherence of the effective error channel. We note that this context is different than that considered in Campbell's convex hull finding algorithm\cite{Campbell2017}: in that algorithm, access to a Solovay-Kitaev-like compiler that can generate approximations to a gate within some error tolerance is assumed. In this work, we suppose that a collection of pre-compiled gates is available to choose from, and we aim to produce an optimal gate given these gates. We will begin by discussing the two broad optimization targets introduced above: i) generator-exact MQGs and ii) Pauli-exact MQGs. We then propose a set of secondary optimization targets that can improve the performance by i) incorporating robustness to drift, ii) targeting low-error-rate solutions, and iii) reducing the number of native unitaries that contribute to the effective channel. 

\subsection{Convex Programs for Constructing Mixed Quantum Gates}
\label{sec:convex_programs}
\subsubsection{Generator-exact MQGs} 
\label{sub:first_order_generators}
\noindent A compelling reason to construct an MQG is to improve the worst-case performance of a quantum gate. The diamond distance bounds this worst-case performance, and so one might assume minimizing it would be a natural optimization target:
\begin{equation}\label{eq:ddmin}
  \begin{split}
    &\underset{\weight_i\geq0, \sum_i\weight_i=1}{\textbf{minimize}: } \hspace{.5cm} \dnorm\left(\sum_i\weight_i\error_i \right)
  \end{split}
\end{equation}
The diamond distance, however, is a non-linear function that in general requires its own convex optimization to compute\cite{watrous2018theory}. This can dramatically slow down iterative optimizers, and so directly minimizing the diamond distance is a computationally burdensome optimization target. However, for quantum computing purposes the error rates are typically quite small, so we can consider a linearized problem, minimizing the diamond distance at first-order in the effective error generator. As discussed around Eq.~\eqref{eq:cancel}, if the origin lies in in the convex hull of the generators, $\0 \in \mathsf{conv}(\{\genmat_i\})$, then we can construct a generator-exact MQG. In \cite{Campbell2017} Campbell presents an iterative algorithm that, given an oracle able to produce approximate unitary operations, will continually generate controls until the geometric constraint is satisfied and then identify the optimal weighting. 

As discussed in the previous section, the geometric condition illustrated in Fig.~\ref{fig:vectorspace} may not be satisfied, due perhaps to stochastic errors in the generators or possibly a situation in which one has access only to $m$ distinct, precomputed gate implementations. In this case, we may simply wish to to identify the MQG with minimal error. A heuristic that is efficiently computable is to minimize the Frobenius norm of the first-order effective error generator:
\begin{equation}\label{eq:frobmin}
  \begin{split}
    &\underset{\weight_i\geq0, \sum_i\weight_i=1}{\textbf{minimize}: } \hspace{.5cm}  \left\vert\left\vert\sum_i \weight_i \genmat_i \right\vert\right\vert_F
  \end{split}
\end{equation}
This optimization problem is equivalent to:
\begin{equation}\label{eq:minimization}
  \begin{split}
    &\underset{\weight_i\geq0, |\weight|_1=1}{\textbf{minimize}: } ||\mathbf{L}\cdot\vec\weight||_2\\
  \end{split}
\end{equation}
Where $\mathbf{L}=\left( \mathsf{col}(\genmat_1) \; \mathsf{col}(\genmat_2) \; \cdots \;  \right)$ is the $d^2\times m$-dimensional matrix of column-stacked error generators and $||\cdot||_2$ is the $\ell_2$ norm. 

The constraints on these optimizations are required to ensure that the weights, $\weight_i$, form a proper probability distribution. Linearly constrained minimization problems with quadratic cost functions are convex and have been proven to be efficiently solvable by, e.g. the ellipsoid method \cite{wright1999numerical, khachiyan}. Many existing convex solver software packages are available that can solve these problems efficiently in practice\cite{cvxpy, cvxpy_rewriting}.

\subsubsection{Pauli-exact MQGs} 
\label{sub:off_diagonals}
As quantum devices grow in size, it is increasingly expensive to simulate their dynamics. Coherent errors impose a particular burden on classical simulators, as the entire quantum state must be tracked continually. Existing work generally approximates channels as Pauli-stochastic channels, and assesses the quality of these approximations\cite{Puzzuoli2014, Magesan2013}. Pauli Frame Randomization adds additional gates into circuits leaving the computation unchanged but randomizing the noise in such a way that it becomes a Pauli-stochastic channel. MQGs offer another path forward. Pauli-exact MQGs can be constructed with vanishing coherent error and without adding any additional gates or compilation time. Building MQGs with diminished coherent error enables the use of much more efficient Pauli-stochastic simulators that are able to model significantly more qubits than their vector-state counterparts. 

In particular, the well-known stabilizer formalism \cite{quant-ph/9807006} for quantum simluation gives a polynomial-time algorithm to compute expectation values of circuits consisting of only Clifford operations. These circuits are useful because they give examples of quantum circuits whose performance can be efficiently assessed using classical simulation. This has motivated significant development of these techniques, including extensions to simulating stabilizer circuits with mixed state inputs\cite{Aaronson2004} and linear combinations of stabilizer circuits\cite{Yoder, Bennink2017}. An interesting property of such simulation techniques is that the cost of the classical simulation is proportional to how non-Clifford the circuit is. Therefore Pauli-exact MQGs have a clear benefit - if the error channels of a circuit can be made to look more like Pauli-stochastic errors, they will be more amenable to efficient classical simulation. 

Like generator-exact MQGs, the optimal weights can be found efficiently by a convex optimization. As above, we assume that we have access to $m$ distinct gate implementations, each of which is associated with its own error matrix, $\errmat_i$. We wish to construct an MQG whose effective error is Pauli-stochastic and is therefore diagonal. This can be done with the following simple minimization:
\begin{equation}\label{eq:minimization2}
  \begin{split}
    &\underset{\weight_i\geq0, \sum_i\weight_1=1}{\textbf{minimize}: } ||\sum_i\weight_i \errmat_i - \mathsf{diag}(\sum_i \weight_i \errmat_i)||_2\\
  \end{split}
\end{equation}
Here $(\mathsf{diag}(\mathcal{A}))_{ij} = \mathcal{A}_{ij}\delta_{ij}$ sets the off-diagonal elements of a matrix to zero. This can be more transparently written as a convex program by constructing the $(d^2-d)\times m$-dimensional matrix, $\mathbf{E}$, whose $i$\textsuperscript{th} column is a vector composed of all of the off-diagonal entries of the error generators $\errmat_i$. \footnote{Exactly \emph{how} the off-diagonal entries are vectorized is unimportant, so long as it is consistent across error matrices.} Eq.~\eqref{eq:minimization2} is then equivalent to 
\begin{equation}\label{eq:minimization3}
  \begin{split}
    &\underset{\weight_i\geq0, |\weight|_1=1}{\textbf{minimize}: } ||\mathbf{E}\cdot\vec\weight||_2
  \end{split}
\end{equation}
The similarity with Eq.~\ref{eq:minimization} is clear and the same methods can be used to solve both. If the solver routine finds a solution for which the cost function is equal to 0, then the resulting error channel will be \textit{exactly} a Pauli channel. In Sec.~\ref{sec:results} we demonstrate the construction of a Pauli-exact MQG, and demonstrate a reduction in the off-diagonal terms of the Pauli transfer matrix by performing a randomized benchmarking experiment.


\subsection{Secondary Objectives}
\noindent 
Often, the cost of generating distinct gate implementations is quite small, so large ensembles of them can be computed rather quickly. One may then have available many more distinct gate implementations than there are parameters in the relevant vector space (see Fig.~\ref{fig:vectorspace}). This can lead to the minimization problems discussed in the previous section being massively under-constrained, yielding a large continuous family of exact solutions. Secondary objectives provide a means for choosing among this family of exact MQGs those with increased performance in a desired area. In this section we present explicit convex programs for such secondary objectives, including adding robustness to drift or model uncertainty, reducing the effective gate error, and minimizing the number of constituent gates utilized by the MQG. We choose to take generator-exact MQGs as our starting point, though the following sections apply equally well to Pauli-exact MQGs with only trivial modifications. 

\subsubsection{Robustness to drift and model uncertainty} 
\label{sub:adding_robustness}
\noindent 
Mixed quantum gates can offer significant performance improvements over bare unitary gates, but their construction requires a good knowledge of the errors experienced by the constituent gates. Often, however, the gates are not perfectly well-characterized. This could be due to simple uncertainty about the parameters of the model used to generate the control solutions, or the system could experience some degree of drift \cite{1907.13608}. In either case, the performance of the MQG will be negatively impacted. We can describe the effect of this uncertainty by first writing the gate explicitly as a function of a vector of model parameters, $\vec\mu$, and their nominal values, $\vec\mu_0$. These model parameters could include magnetic field strengths, laser or microwave intensities, coupling strengths, etc. If the parameters are close to their nominal values, then we can write a gate's error generator as a Taylor expansion in the deviation vector, $\vec\delta = \vec\mu-\vec\mu_0$:
\begin{equation}
	\label{eq:sensitivity_expansion}
	\genmat_i(\vec \delta) = \genmat_i(\0) + \sum_k \delta_k \,\genmat_{i,k}(\0) + \order{\delta^2}.
\end{equation} 
In the above expression we have used the comma derivative, 
\begin{equation}
	\genmat_{i,k}(\vec \delta_0) = \frac{\partial}{\partial \delta_k} \genmat_i(\vec\delta) \vert_{\vec \delta=\delta_0}.
\end{equation}

Combining Eq.~\eqref{eq:sensitivity_expansion} with Eq.~\eqref{eq:generator} and taking both Taylor expansions to first order, we have:
\begin{align}
\notag
	\errmat_{\rm eff} =
		 \identmat_d &+ \sum_i \weight_i \left(\genmat_i(\0) + \sum_k \delta_k \,\genmat_{i,k}(\0) \right) \\
		 &+ \order{\genmat_i^2 + \genmat_i \delta + \delta^2}.
\end{align}
We would like to choose the weights $\vec\weight$ so that the effective error generator vanishes to first order. We can cast this problem as a convex optimization problem, much as we did in Sec.~\ref{sub:first_order_generators} for synthesizing generator-exact MQGs. Following Eq.~\eqref{eq:minimization}, we have,
\begin{equation}
  \begin{split}
    &\underset{\weight_i\geq0, |\weight|_1=1}{\textbf{minimize}: } ||\mathbf{L}\cdot\vec\weight||_2 + \sum_k||\mathbf{L}_{,k}\cdot\vec\weight||_2\\
  \end{split}
\end{equation}
Where $\mathbf{L}_{,k}=\left( \mathsf{col}(\genmat_{1,k}(\0)) \; \mathsf{col}(\genmat_{2,k}(\0)) \; \cdots \;  \right)$ is the $d^2\times m$-dimensional matrix of column-stacked derivatives of the error generators with respect to the deviation $\delta_k$.

\begin{figure}
  \centering
  \includegraphics[width=\columnwidth]{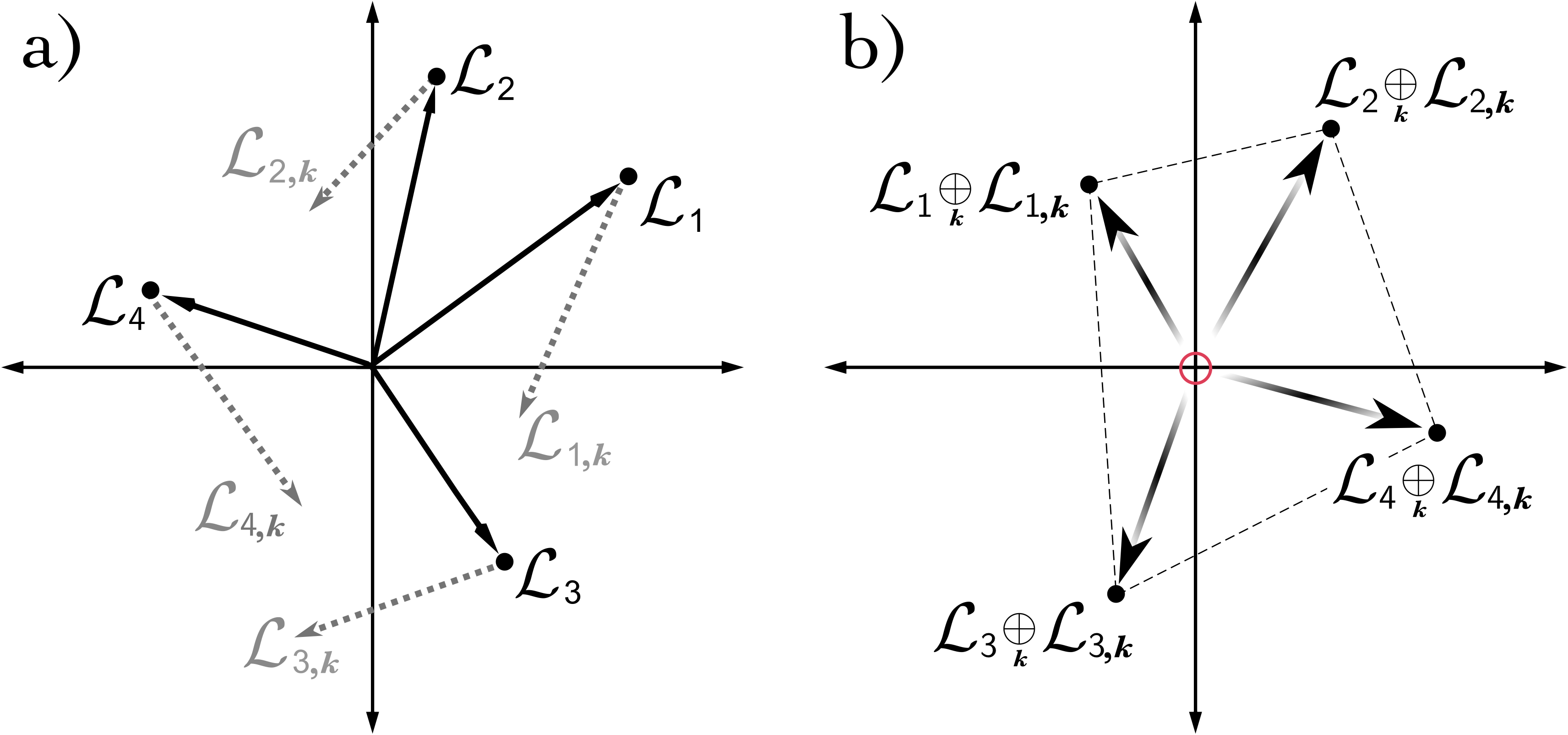}
  \caption{\textbf{Geometric condition for robustness:}   A target unitary gate can be implemented a number of ways, each with a different effective error generator, $\genmat_i$. a) The error generators are shown as elements of a vector space (solid, black arrows). Also shown are their derivatives $\genmat_{i,k}$ (dashed, grey arrows) with respect to a model parameter, $\delta_k$. As this parameter drifts, the generators may no longer cancel. b) To construct a robust, generator-exact MQG, there must exist a set of weights, $\{\weight_i\}$, so that $\sum_i \weight_i \genmat_i = \sum_i \weight_i \genmat_{i,k} = 0$. This is equivalent to demanding that the origin lie in the convex hull $\mathsf{conv}(\{\genmat_i \bigoplus_k \genmat_{i,k} \}_i)$. The gradient arrows indicate that this space is higher-dimensional. }
  \label{fig:vectorspace2}
\end{figure}

We can additionally generalize the geometric criterion for guaranteeing the existence of generator-exact MQGs. To construct a robust, generator-exact MQG, there must exist a set of weights, $\{\weight_i\}$, so that $\sum_i \weight_i \genmat_i = \sum_i \weight_i \genmat_{i,k} = 0$. This is equivalent to demanding the origin lie in the convex hull of the generators augmented with their derivatives:
\begin{equation}
\0 \in \mathsf{conv}\left( \left\{
\genmat_i \bigoplus_k \genmat_{i,k}
\right\}_i	 \right).
\end{equation}
We illustrate this in Fig.~\ref{fig:vectorspace2}.

We call a gate constructed in this way a \emph{robust} generator-exact MQG. These techniques can be easily adapted to construct robust Pauli-exact MQGs, where vectors of the off-diagonal elements of the error maps take the role of the error generators used above.

\subsubsection{Improving the average gate fidelity}
\label{sec:norm}
\noindent While the geometric constraint is sufficient for suppressing the diamond norm to first order relative to the \textit{worst} controls in the collection, it does not preferentially select the best controls possible.  As an example, consider four erroneous implementations of a Pauli $Z$ gate: $Z_{-2\theta}$, $Z_{-\theta}$, $Z_{\theta}$, and $Z_{2\theta}$, where the subscript indicates the magnitude of the $Z$-rotation error. An MQG consisting of equally weighted $Z_{-2\theta}$ and $Z_{2\theta}$ is generator-exact, as is $Z_{-\theta}$ and $Z_{\theta}$. But the error rate\footnote{The diamond distance is equal to the AGI here.} of the first is $~4\theta^2$, while the second achieves an error rate of $\theta^2$. 

To incentive the inclusion of controls with smaller error, we may instead minimize the average gate infidelity of the MQG, subject to the condition that any solution we find be generator-exact:

\begin{equation}\label{eq:agi_minimization_constrained}
  \begin{split}
    &\underset{\weight_i\geq0, |\weight|_1=1}{\textbf{minimize}: } \sum_i \weight_i\; \AGI(\errmat_i)\\
    & \textbf{subject to}: ||\mathbf{L}\cdot\vec\weight||_2 = 0 
  \end{split}
\end{equation}

If we do not expect to have a space of generator-exact solutions available, then the constraint above will never be satisfied and the minimization will fail. This could occur if there are simply not enough controls to satisfy the generator-exact criterion, or if the error generators have stochastic components. In such a case, we can choose a weighting parameter, $\eta$, to explicitly balance the two competing objectives:
\begin{equation}\label{eq:agi_minimization_weighted}
  \begin{split}
    &\underset{\weight_i\geq0, |\weight|_1=1}{\textbf{minimize}: } ||\mathbf{L}\cdot\vec\weight||_2  + \eta \sum_i \weight_i\; \AGI(\errmat_i)
  \end{split}
\end{equation}
Both optimizations are convex and efficiently solvable\cite{Shah2001}. 


\subsubsection{Sparsity Constraints}
\label{sec:sparsity}
\noindent
As a practical consideration, we would also like to regularize our objective function to promote sparse weightings, resulting in MQGs composed of only a small number of individual gate implementations. Control electronics often have a limited amount of waveform memory, and thus it is important to be able to construct MQGs with as few constituent gates as possible. In many machine learning contexts, a technique called lasso regularization \cite{tibshirani1996regression} can be used to enforce this sparsity in solutions. 

Rather than penalizing the $\ell_0$ norm of the solution, which is not convex, lasso regularization penalizes the $\ell_1$ norm, which is a convex relaxation of the $\ell_0$ norm. However this is insufficient for our purposes, as we already require $\weight$ to be a valid probability distribution, which constrains the $\ell_1$ norm to be equal to $1$.

The problem of enforcing sparsity in such situations has been considered in \cite{NIPS2012_4504} and can be expressed via another convex program that extends Eq.~\eqref{eq:minimization}:
\begin{equation}\label{eq:minimization_sparsity}
\begin{split}
&\underset{m\in[1..N]}{\textbf{minimize:}}\\
&\ \ \bunderbrace{\textbf{\hspace{.2cm}minimize:\hspace{.2cm}} }{\hspace{.1cm}\weight_i\geq0, |\weight|_1=1, t\geq0} ||\mathbf{L}\cdot\vec\weight||_2 + t\\
&\ \ \hspace{.2cm}\textbf{subject to: } \weight_m > \frac{\lambda}{t}.
\end{split}
\end{equation}

Here, $\lambda \geq 0$ is a tunable parameter that can be used to control the degree of sparsity in the solution. If $\lambda$ is set to zero, the inner program reduces to Eq.~\eqref{eq:minimization} and the outer minimization is redundant. If $\lambda$ is very large, each inner minimization problem is minimized for an MQG with a single control, and the outer minimization simply selects the control which has the error generator of smallest norm. While this gives a maximally sparse solution, it doesn't allow for any cancellation of coherent errors.

In general, increasing $\lambda$ decreases the number of controls with large weights, and by removing controls with weights below a threshold probability we can increase the sparsity of the solution. Truncating in this way will introduce error in the generator-exactness of the solution. For larger $\lambda$ the required threshold probability decreases, and so the error introduced from truncation can be made smaller. However, larger $\lambda$ will also decrease the likelihood of the optimizer returning a generator-exact solution, and so in this way, tuning $\lambda$ allows us to trade between different sources of error when constructing sparse MQGs. An explicit example of the usage of this optimization constraint is given in Sec.~\ref{sub:experimental}, and shown in Fig.~\ref{fig:sparsity}, where the optimizer chooses just $10$ of $750$ controls.

In the following sections we discuss both experimental and numerical implementations of MQGs, leveraging each of these secondary optimizations.

\section{Results} 
\label{sec:results}


\subsection{Experimental implementation } 
\label{sub:experimental}
\noindent We implemented our methods using the 19Q-Acorn superconducting transmon processor at Rigetti Computing. Comprehensive characterization of this device can be found in \cite{1712.05771}. While this device consisted of 20 qubits with fixed couplings, we used only qubit \#8 for our demonstration.

To construct an MQG, there needs to be several different realizations of the desired gate available. To see improved performance, these realizations need to be imperfect in different ways, from drift, miscalibration, or other sources of error. Because the qubit used in our demonstration was coherence-limited, and therefore did not suffer from miscalibration errors, we intentionally introduced coherent errors to showcase our technique.  Controls were derived from a single, 10-sample, 50ns Gaussian $X_{\pi/2}$ pulse calibrated via Rabi oscillation. Four intentionally miscalibrated Gaussian pulses, labeled $\mathsf{Pulse1}$ through $\mathsf{Pulse4}$, were derived by scaling the amplitude of the calibrated $X_{\pi/2}$ pulse by a factor $S \in \{1.064, 1.039,0.937, 0.912\}$, respectively. We model their effect on the qubit state as perfectly unitary rotations, $X_{A \pi/2} = \exp{(-i A \pi/2 \sigma_x)}$.

\begin{figure}
\includegraphics[scale=.5]{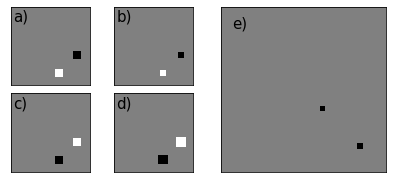}
\caption{Hinton diagrams representing the error generators for a) $\mathsf{Pulse1}$, b) $\mathsf{Pulse2}$, c) $\mathsf{Pulse3}$, d) $\mathsf{Pulse4}$ and e) a Pauli-exact MQG. White boxes correspond to positive values, black  boxes correspond to negative values, and the size of the box corresponds to the magnitude of each entry. The four constituent pulses can be seen to have coherent errors from the off-diagonal components of their error generators. The Pauli-exact MQG has been constructed to have minimal coherent error and therefore has only diagonal elements.}   
\label{fig:ptms}
\end{figure}

 These gates are then used as in Sec.~\ref{sub:off_diagonals} to construct a Pauli-exact MQG, which samples from the miscalibrated waveforms with probabilities $P\in \{0.307, 0.283, 0.211, 0.199\}$ respectively. Using these probabilities and pulse amplitudes we have plotted the corresponding error generators in Fig.~\ref{fig:ptms}. $\mathsf{Pulse1}$ through $\mathsf{Pulse4}$ show clear coherent errors as off-diagonal elements while the MQG exhibits a clear suppression of these off-diagonal elements.

We characterize the performance of the MQG using a series of randomized benchmarking (RB) experiments\cite{Magesan2011}: one for each of the miscalibrated pulses, one using the calibrated pulse, and using the MQG. In each case, the Clifford operations were decomposed into $X_{\pi/2}$ gates, which were implemented directly using the relevant waveform, and $Y_{\pi/2}$ gates, which were implemented by phase-shifting the relevant waveform $\pi/2$ radians. All circuits were precompiled on the host computer and sent to be executed on the control electronics, rather than using the FPGA or lower-level firmware to sample from $\mathsf{Pulse1}$ through $\mathsf{Pulse4}$ at run-time.

An RB experiment consisted of sampling and running $10$ random Clifford sequences at each of lengths $L\in\{2, 4, 8, 16, 32, 64\}$. Each sequence was repeated $1000$ times. We emphasize that, when benchmarking the MQG, each sequence repetition used a new, randomly sampled sequence of waveforms consistent with its definition. Results of the standard RB analysis are discussed in Table.~\ref{tabl:rb}. The MQG is seen to perform nearly as well as the calibrated pulse, and better than any of the constituent pulses individually. 

\setlength{\tabcolsep}{0.5em} 
\renewcommand{\arraystretch}{1.2}
\begin{table}[h]
	\centering
	\begin{tabular}{@{}ccc@{}}
		\hline
		Pulse name & Rotation error & RB error rate\\
		\hline
		$\mathsf{Pulse1}$ 	& 6.4\% 	& 1.1\% \\
		$\mathsf{Pulse2}$ 	& 3.9\% 	& 0.9\% \\
		$\mathsf{Pulse3}$ 	& --\,6.3\% 	& 1.1\% \\
		$\mathsf{Pulse4}$ 	& --\,8.8\% 	& 1.5\% \\
		Calibrated 			& $\cdot$  	& 0.7\% \\
		MQG 				& $\cdot$ 	& 0.8\% \\
		\hline
	\end{tabular}
	\caption{\textbf{Randomized benchmarking results:} The MQG outperforms all of the individual miscalibrated gates of which it is composed, achieving an RB number nearly as low as the carefully calibrated pulse. }
\label{tabl:rb}
\end{table}

\begin{figure}[t]
  \centering
  \includegraphics[width=\columnwidth]{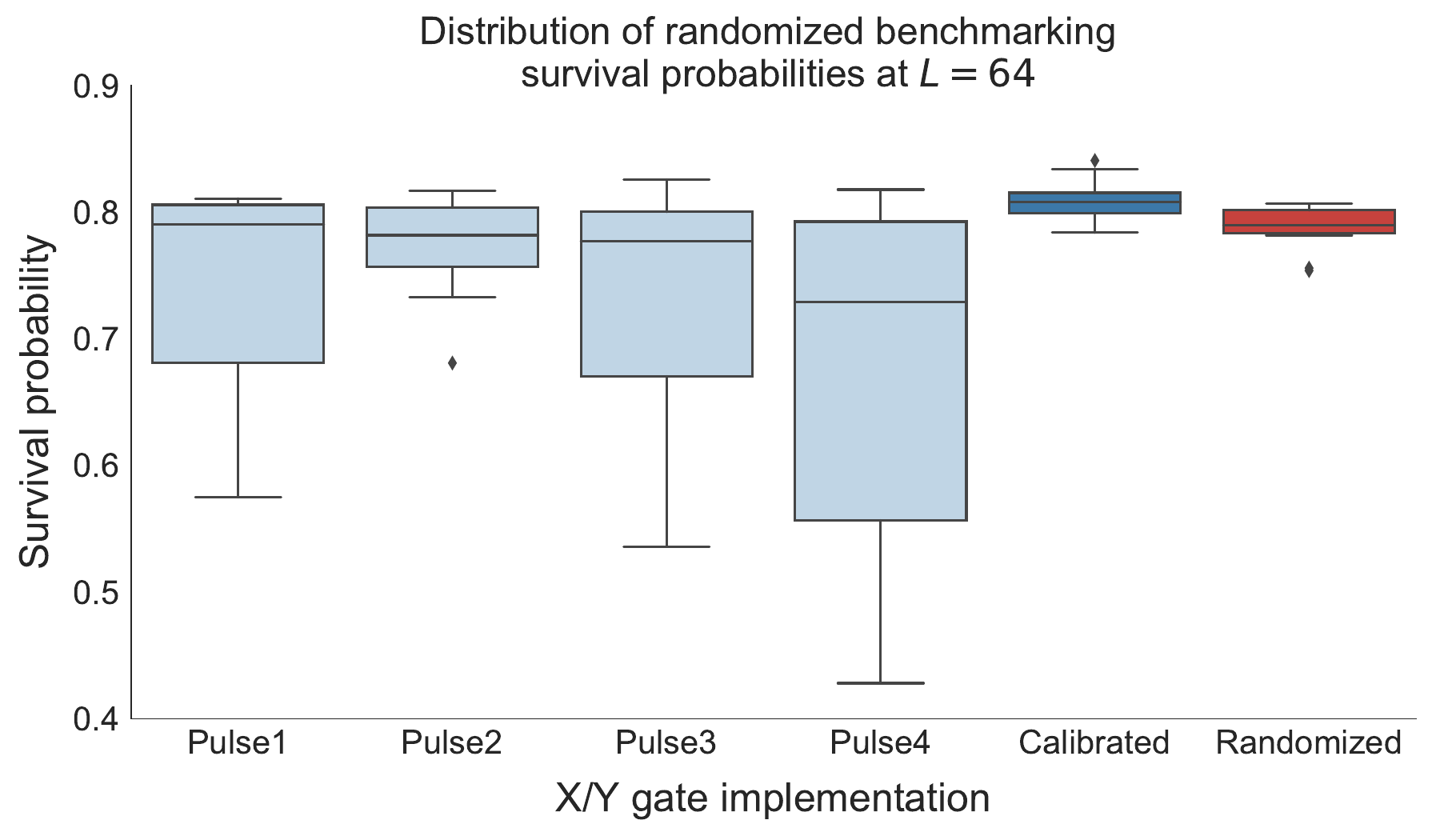}
  \caption{\textbf{Evidence of reduced coherent error:}. Shown here is the distribution of randomized benchmarking survival probabilities at $L=64$ for each of six native gate implementations. The first four boxes (light blue) are the results from four intentionally miscalibrated ${\pi/2}$ rotations. The coherent noise present in these implementations leads to large variance in the survival probability over sequences. The fifth (dark blue) box illustrates the survival probability using a highly-tuned gate implementation. It displays improved average survival probability as well as reduced variance. The final box (dark red) illustrates the distribution for a randomized MQG composed of $\mathsf{Pulse1}$ through $\mathsf{Pulse4}$. It performs comparably to the highly-calibrated implementation in both average survival probability and variance over random sequences. The reduced variance of the MQG is a tell-tale sign of reduced coherent error in the effective channel\cite{Ball2016}.}
  \label{fig:rb}
\end{figure}

By minimizing the off-diagonal elements of the process matrix, we expect the resulting MQG to display suppressed coherent error. To see that this is in fact the case, we can inspect the variance of the RB survival probabilities for the MQG relative to the miscalibrated pulses. As discussed in \cite{Ball2016}, coherent errors will tend to broaden the distribution of RB survival probabilities over sequences at each length, generally manifesting as a long-tailed gamma distribution. Stochastic noise, such as depolarizing noise, will yield comparatively narrow, Gaussian-distributed success probabilities. In Fig.~\ref{fig:rb}, we plot the experimentally observed distribution of survival probabilities at sequence length $64$ for each of the benchmarked gate sets. We see that the intentionally miscalibrated controls in our RB experiment have long tails, consistent with coherent noise, while the calibrated and randomized implementations are both significantly less dispersed, consistent with stochastic errors. This experiment therefore provides compelling evidence that the MQG suffers considerably less coherent error than any of the miscalibrated gates from which it was constructed.

We note that, while this example is somewhat contrived (the calibrated gate is clearly the best), it nonetheless succeeds in demonstrating that the MQGs are capable of outperforming the individual gates of which they are composed. In future work, we hope to apply this technique to two-qubit gates, which are significantly more difficult to tune up in general and are more likely to possess lingering coherent errors. Tomography or modeling can then be used to construct the process matrix estimates required to build a mixed gate. 


\subsection{Numerical implementations}
\label{sec:numerical_results}
\noindent In the following numerical results, we use the various methods in Section \ref{sec:mixed_unitary_processes} to build a range of one- and two-qubit MQGs for model systems. We analyze the resulting mixed gates for a range of Hamiltonian parameters, demonstrating both a reduction in diamond distance and an improved robustness to model uncertainty. 

\subsubsection{Single-qubit example} 
\label{sub:one_qubit}
We consider the following dimensionless model for a single qubit subject to frequency drift: 
\begin{equation}\label{eq:1Qham}
  H(\delta, \epsilon, t) = \epsilon\sigma_z + (1 + \delta)(c_x(t)\sigma_x + c_y(t)\sigma_y)
\end{equation}
where $\epsilon$ corresponds to fluctuations in qubit frequency and $\delta$ corresponds to fluctuations in the control field.

\begin{figure}
\includegraphics[width=\columnwidth]{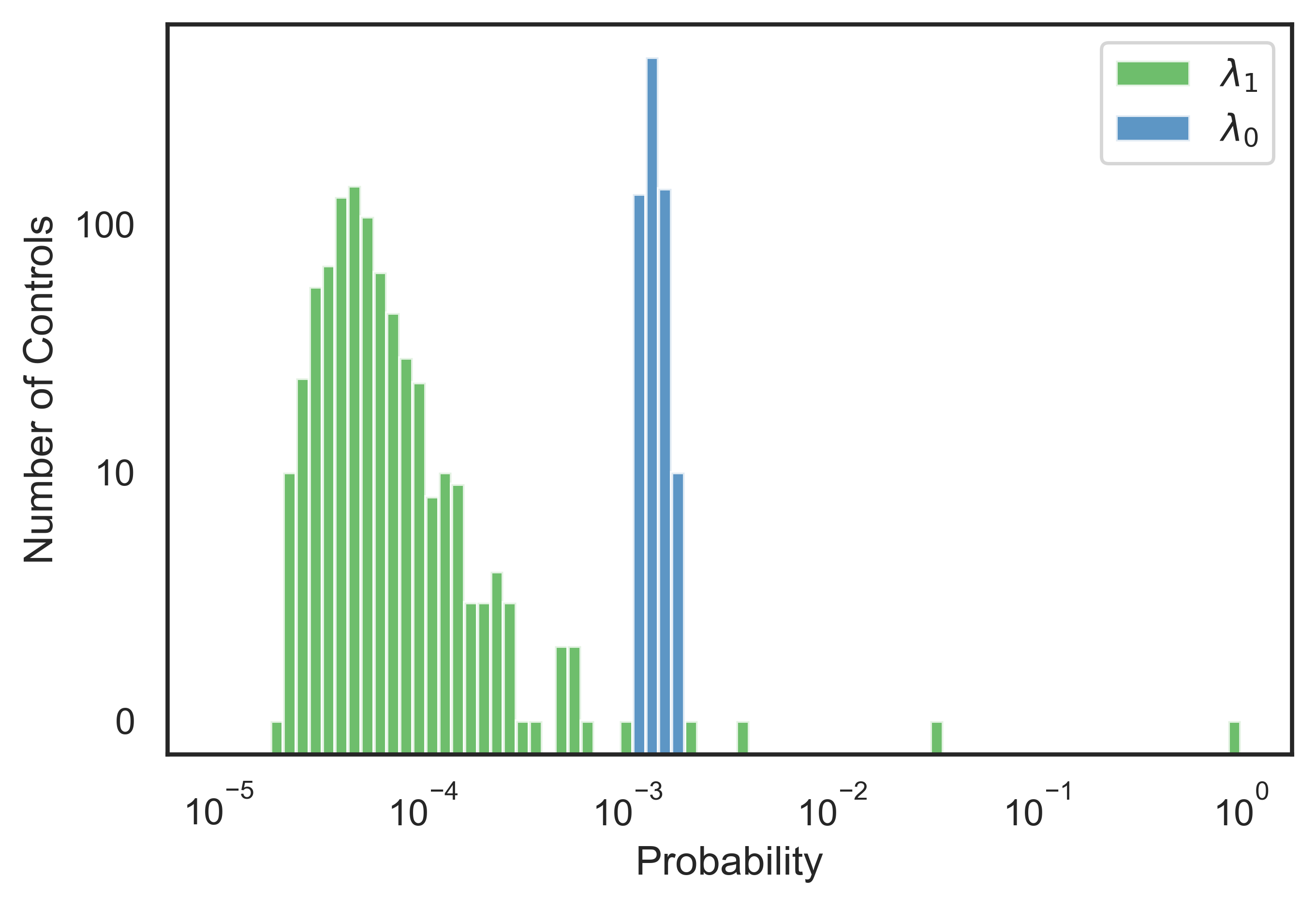}
\caption{Sparsity of MQGs as a function of the sparsity parameter $\lambda$, with $\lambda_0=0$ and $\lambda_1=0.0005$. As the parameter is increased from zero, the number of controls with low probability increases (seen on the plot as the probability mass moving to the left). For $\lambda_1$ virtually all of the 750 controls have vanishingly small probability, allowing for the restriction of the MQG to a small number of controls.}
\label{fig:sparsity}
\end{figure}
To generate the initial controls, we use the GRAPE algorithm\cite{Khaneja2005} with N=25 steps and total evolution time of $\pi$ to generate 750 candidate controls. We give a detailed discussion of GRAPE in the  appendix (\ref{sub:grape}). Using the results from in Section \ref{sec:sparsity} we generated MQGs that sample from only 10 of these controls. Fig.~\ref{fig:sparsity} shows solutions to Eq.~\eqref{eq:minimization_sparsity} for different choices of  $\lambda$. As $\lambda$ increases, most of the weights in the optimal solution become smaller, resulting in a sparser MQG. 

In our implementation of the GRAPE algorithm, we use the performance function presented in \cite{Khaneja2005}, and average over different values of $\delta$ and $\epsilon$ using Gaussian quadrature when computing the gradient, so that we find controls that are naturally robust. The standard deviations considered for all parameters in our numerical experiments were fixed to $\sigma=.001$. Finally, we assume that the errors on $\sigma_x$ and $\sigma_y$ are perfectly correlated, as in our experimental implementation. We note that advanced quantum control protocols may provide an even more principled approach to control synthesis. DMORPH \cite{dominy2008exploring}, for example, explores continuous families of controls on fidelity level sets, thereby enabling further optimizations against secondary criteria, such as the duration of the control pulse or robustness to drift. 

\begin{figure}[h]
  \centering
  \includegraphics[width=\columnwidth]{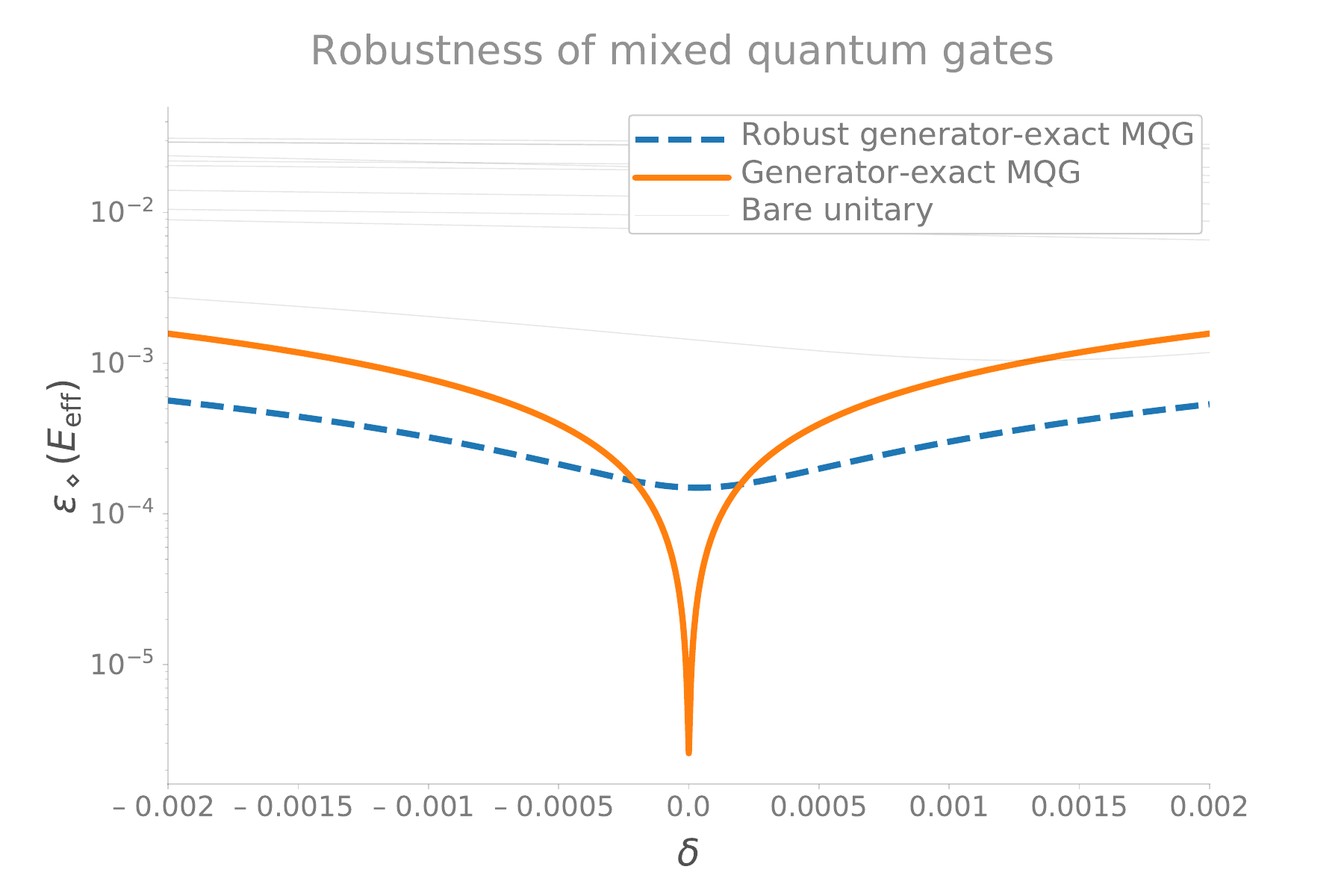}
  \caption{\textbf{One-qubit robust generator-exact MQG} Numerical results showing the diamond distance errors for a generator-exact MQG (dashed blue), a robust generator-exact MQG (solid orange), and a large set of bare unitary controls (thin solid grey) as a function of a the fractional amplitude error. The target is an $X_{\pi/2}$ gate on a single qubit. The generator-exact MQG dramatically outperforms both the robust generator-exact and any of the bare unitaries at $\delta=0$. However, even a relatively small amplitude error leads to a regime in which the robust generator-exact will yield the better gate. Similar results are obtained for drift in the qubit frequency.}
  \label{fig:YMQG}
\end{figure}

Using controls generated in this way, the MQGs produced for $X_{\pi/2}$ and $Y_{\pi/2}$ are qualitatively similar, with the results for $X_{\pi/2}$ shown in Fig. \ref{fig:YMQG}. By imposing the penalty from Section \ref{sec:norm}, we sought to ensure that the algorithm preferentially selected controls with smaller errors. Adding this constraint increased the performance of the generator-exact MQG by nearly an order of magnitude at the origin, and produced a robust generator-exact MQG whose performance is an order of magnitude better than the generator-exact MQG away from the origin. Imposing this constraint allows us to trade off flatness for performance. This shows that through adding constraints to our optimization routine, we can make the MQG practically useful. 

\subsubsection{Two-qubit example} 
\label{sub:two_qubit}
In our two-qubit example we consider a resonant exchange interaction, similar to that in \cite{McKay2016}:
\begin{equation} \label{eq:2Qham}
\begin{split}
H(\vec{\delta}, \vec{\epsilon}, t) = &\sum_{j=1}^2(\epsilon_j\sigma_z^j + (1 + \delta_j)(c_x^j(t)\sigma_x^j + c_y^j(t)\sigma_y^j)) \\
&+ \frac{1}{10}(XX + YY)
\end{split}
\end{equation}

In this example it was infeasible to use GRAPE to return non-trivial solutions, as the algorithm would tend to return similar solutions for a wide variety of initial conditions. Instead we manually selected piecewise constant echoing sequences with 500 steps and total evolution time of $\frac{5\pi}{2}$. In particular, we considered $RX(\pi)$, $RX(-\pi)$, $RY(\pi)$ and $RY(-\pi)$ bang-bang sequences \cite{bangbang}, consisting of all combinations of simultaneous $\pi$ pulses activated at multiples of $8$ steps from the beginning of the controls, and the same multiple of $8$ steps prior to the end of the controls. To give the control family a variety of RF errors, we also added uniformly distributed amplitude errors to each $\pi$ pulse, between $-.25$\% and $.25$\%.

\begin{figure}
  \centering
  \includegraphics[width=\columnwidth]{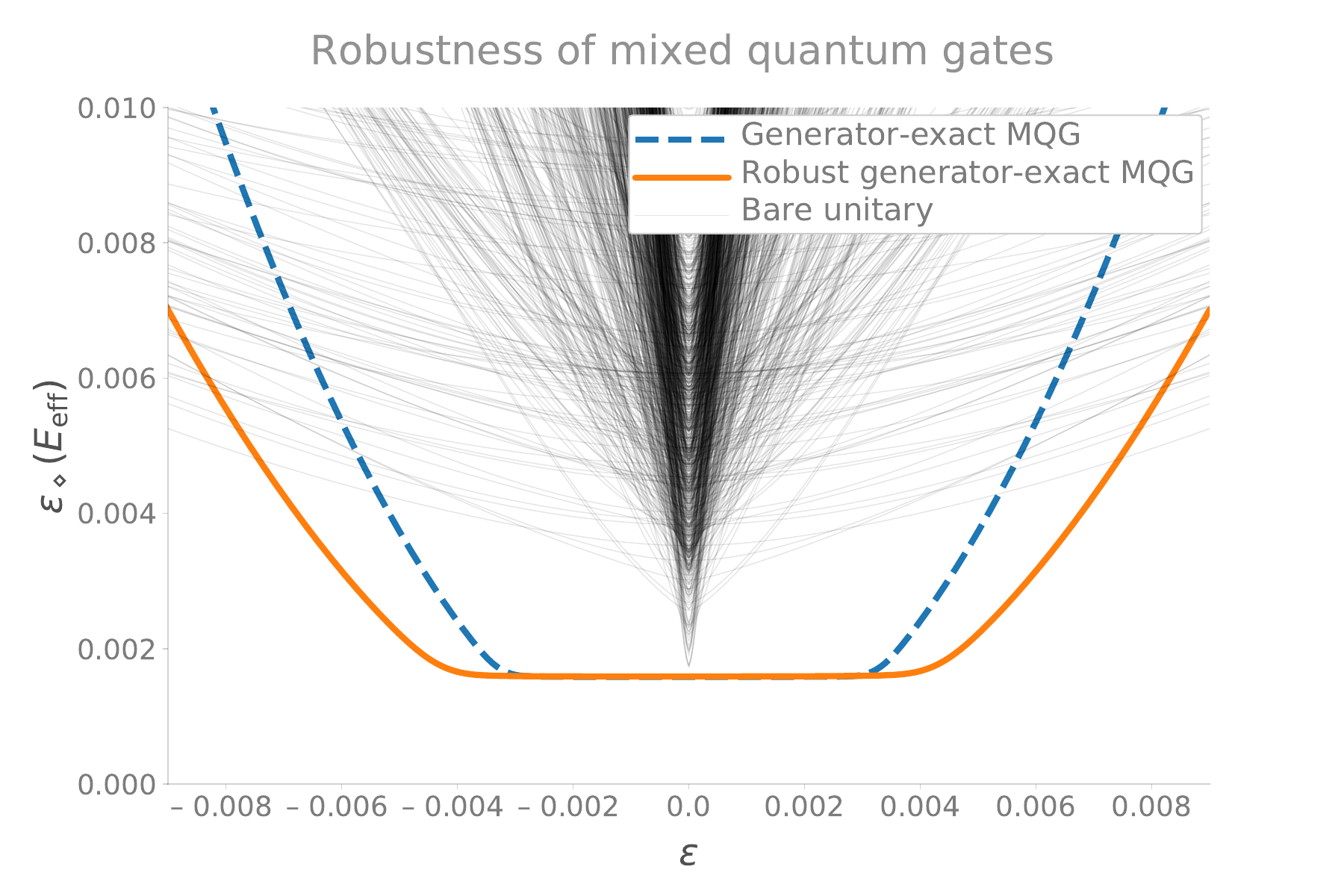}
  \caption{\textbf{Two-qubit robust generator-exact MQG} Numerical results showing the diamond distance errors for a generator-exact MQG (dashed blue), a robust generator-exact MQG (solid orange), and a large set of bare unitary controls (thin solid grey) as a function of a spurious detuning (quantified by the ratio of the detuning to the maximum control amplitude). It is intended to model a pair of qubits interacting via resonant exchange. Both MQGs can be seen to outperform the bare unitary control over a wide range of detunings. The robust generator-exact  MQG always outperforms the generator-exact MQG, and both MQGs outperform any of the bare unitary controls over a wide range of detunings. Similar results are obtained for drift in the qubit control amplitude.}
  \label{fig:2MQG}
\end{figure}

In this example, we find more modest improvements to performance, as shown in Fig. \ref{fig:2MQG}. There are now four free parameters to optimize over, and the uncontrolled entangling interaction means that there is little room for variation in the controls. Nonetheless, using an MQG improves performance over any of the constituent controls over a wide-range of detunings, and for all values of the detunings we see that the robust generator-exact MQG performs as well or better than generator-exact MQG.

\section{Conclusion and Future Work}
We have shown numerically that using MQGs can reduce coherent error by more than an order of magnitude in diamond norm, over a wide range of quasi-static values of noise. In addition, we have demonstrated that these approximate controls can be generated through optimal control (GRAPE), and that the minimization problem is tractable. By implementing four miscalibrated $X_{\frac{\pi}{2}}$ rotations on a superconducting qubit and drawing from them with optimized weights, we see that coherent over- and under- rotations can be averaged away.

Future directions for this work include demonstrating the routine experimentally on a two-qubit gate, moving the random gate selection from a pre-compilation step to runtime logic onboard the control electronics, investigating other control generation algorithms such as CRAB \cite{Caneva2011} and GOAT\cite{Machnes2018}, and using more sophisticated benchmarking routines such as GST\cite{BlumeKohout2017} to quantitatively investigate the performance of our method.

Additionally while this optimal control in this work has been \textit{ex situ}, an interesting approach would be to instead use \textit{in situ} techniques \cite{Wu2018, Kelly2014, Ferrie2015} to generate the controls. While performing a complete optimization in this way would require full process tomography, one could instead use partial tomography. By selecting pre-- and post --rotations that correspond to measuring Pauli-moments of interest in the Hamiltonian, such as unwanted $Z\otimes Z$ crosstalk terms, one could perform optimization over fewer parameters to produce MQGs that suppress specific errors.


\section{Acknowledgements}
\label{sec:acknowledgements}
This material was funded in part by the U.S. Department of Energy, Office of Science, Office of Advanced Scientific Computing Research Quantum Testbed Program, as well as by Sandia National Laboratories' Laboratory Directed Research and Development program. Sandia National Laboratories is a multimission laboratory managed and operated by National Technology and Engineering Solutions of Sandia, LLC, a wholly owned subsidiary of Honeywell International, Inc., for the U.S. Department of Energy's National Nuclear Security Administration under contract DE-NA0003525.

\bibliography{mixed_unitaries.bib}

\section{Appendix}
\label{sec:appendix}

\subsection{Proof of diamond distance inequality}
\label{sub:diamond_distance_inequality}
Here we prove the claim of Eq.~\eqref{eq:diamond_ineq} that:
\begin{equation}
	\dnorm\!\left(\error_{\rm eff}\right) \le \sum \weight_i \, \dnorm(\error_i).
\end{equation}
The effective error channel for a mixed quantum gate is $\error_{\rm{eff}} = \sum \weight_i\,\error_i$, where $\error_i$ are the error channels for the component gates. The diamond distance to the identity of the effective error channel is:
\begin{align}
	\dnorm \left(\error_{\rm eff}\right)
		&= \frac{1}{2} \sup_\rho \vert \vert (\ident_d\otimes \ident_d)(\rho) 
										  - (\error_{\rm eff} \otimes \ident_d)(\rho) \vert\vert_1\\
		&= \frac{1}{2} \sup_\rho \vert \vert \sum \weight_i\,((\ident_d
										  - \error_i) \otimes \ident_d)(\rho) \vert\vert_1
\end{align}
For qubits, the space of density matrices is compact, so the supremum is achievable. Call a state that achieves the supremum $\rho^*$. Then 
\begin{align}
	\dnorm \left(\error_{\rm eff}\right)
		&= \frac{1}{2} \vert \vert \sum \weight_i\,((\ident_d
										  - \error_i) \otimes \ident_d)(\rho^*) \vert\vert_1 \\
		&= \frac{1}{2} \vert \vert \sum \weight_i\, \rho^*_i \vert\vert_1,
\end{align}
where we have defined $\rho^*_i = ((\ident_d - \error_i) \otimes \ident_d)(\rho^*)$. The nuclear norm above is equal to the sum of the singular values of $\sum \weight_i \rho_i$. Using the Ky Fan singular value inequality \cite{fan1951maximum} , we have 
\begin{align}
	\dnorm \left(\error_{\rm eff}\right)
		&\le \frac{1}{2} \sum_i \weight_i \vert \vert \rho^*_i \vert\vert_1 \\
		&\le \sum_i \weight_i\, \dnorm(\error_i)
\end{align}
The second inequality above follows because $\rho^*$ defines an explicit lower bound for the diamond distance for each of the component error maps. 
\subsection{GRAPE}\label{sub:grape}
In this paper, we use techniques from optimal control theory to generate the control pulses that comprise the MQGs. Optimal control theory is a diverse field, and many techniques exist for generating control pulses. In this work, we utilize the GRadient Ascent Pulse Engineering  (GRAPE) algorithm \cite{Khaneja2005, Caneva2011, Machnes2018}. First described in \cite{Khaneja2005}, GRAPE is a technique for finding piecewise constant control sequences that approximate a desired unitary, $U_T$, given a Hamiltonian with controlled and uncontrolled terms. Defining the uncontrolled Hamiltonian as $H_0$, the control Hamiltonians as $H_{i\neq 0}$, and the \textit{control matrix} $c_{ij}$ as containing the control amplitude associated with the $i^{th}$ time step and the $j^{th}$ Hamiltonian, we can write the unitary for any timestep of evolution as:
\begin{equation}\label{eq:3}
  U_i = \exp\{-i\Delta t(H_0 + \sum_{j=1}^{n}c_{ij}H_{j}\}
\end{equation}
Typically, the error between the  approximate unitary $U=\prod_iU_{i=1}^n$ and the target unitary $U_T$ is quantified using the cost function $J(U) = \mathrm{Re}[Tr\{U_T^{\dagger}U\}]$. This cost function is proportional to the fidelity between unitary operators \cite{Cabrera2011-ji}. 

To optimize this cost function we can perform the following standard update loop for some threshold value $\varepsilon > 0$ and step size $\delta > 0$:
\begin{algorithm}[H]
\floatname{algorithm}
  \caption{\textsc{\textbf{Gradient Ascent}}}
  \begin{algorithmic}
    \While{$J(U) < (1-\varepsilon$)}
    \State $c_{ij} \rightarrow c_{ij} + \delta\frac{\partial J(U)}{\partial c_{ij}}$
    \For{$1 \leq i \leq n$}
    \State $U_i \rightarrow \exp\{-i\Delta t(H_0 + \sum_{i=0}^{n}c_{ij}H_j)\}$
    \EndFor
    \State $U \rightarrow \prod_{i=1}^nU_i$
    \EndWhile
  \end{algorithmic}
\end{algorithm}

In general these gradients can be computed by propagating partial derivatives of the cost function with respect to control parameters through each timestep of the  via the chain rule. However, in \cite{Khaneja2005} Khaneja et al. derive an update formula that is correct to first order, and efficient to compute. In particular one can show that:
\begin{equation}\label{eq:update}
  \begin{split}
\frac{\partial J(U)}{\partial u_{ij}} = -2Re\{\braket{{U_{j+1}^{\dagger}...U_N^{\dagger} U_T}}{i\Delta tH_jU_j...U_1}\\
\braket{U_j...U_1}{U_{j+1}^{\dagger}...U_N^{\dagger} U_T}\} +  \mathcal{O}(\Delta t^2)
  \end{split}
\end{equation}

Because we are trying to generate high-order MQGs, we want to generate controls that perform well even in the presence of drift. To do this, we modified the gradient in GRAPE to instead be:
\begin{align}\label{quadrature}
\frac{\partial \tilde J(U)}{\partial u_{ij}} =
\int p(\vec{\delta})\frac{\partial J(U(\vec{\delta}))}{\partial u_{ij}} d\vec{\delta}
\end{align}
with $p(\vec{\delta})$ Gaussian distributed, as has been done in previous works \cite{Goerz2014} to ensure that the optimal control results are robust over a wide range of errors. To make this averaging tractable, we approximate this integral using Gaussian quadrature, approximating the cost function as a low order polynomial\cite{abramowiz1972handbook} (in particular degree 5). Note that despite updating the gradient, we did not change the cost function. Incorporation of robustness information in the gradient descent step effectively prioritized robust solutions, but did not require changes to the optimization target. This is not a fundamental requirement of the routine, but rather was done to make finding solutions with $J(U) < 1 - \epsilon$  less computationally taxing.
\end{document}